\documentclass[useAMS,usenatbib]{mn2e}
\usepackage{aas_macros}
\usepackage{graphicx}
\usepackage{amssymb}

\begin{document}

\title[Detection of population III stars with the JWST]{Detection of isolated population III stars with the James~Webb~Space~Telescope}
\author[Rydberg et al.]{Claes-Erik Rydberg$^{1}$\thanks{E-mail:
claes-erik.rydberg@astro.su.se}, Erik Zackrisson$^{1}$, Peter Lundqvist$^{1}$ and Pat Scott$^{2}$\\
$^{1}$Department of Astronomy, Stockholm University, Oscar Klein Center, AlbaNova, Stockholm SE-106 91, Sweden\\
$^{2}$Department of Physics, McGill University, Montreal, QC H3A 2T8, Canada}

\date{Accepted 2012 December 18. Received 2012 December 17; in original form 2012 June 1}

\maketitle

\begin{abstract}

The first population III stars are predicted to form in minihalos at $z \approx 10$--30. The {\it James Webb Space Telescope (JWST)}, tentatively scheduled for launch in 2018, will probably be able to detect some of the first galaxies, but whether it will also be able to detect the first stars remains more doubtful. Here, we explore the prospects of detecting an isolated population III star or a small cluster of population III stars down to $z=2$ in either lensed or unlensed fields. Our calculations are based on realistic stellar atmospheres and take into account the potential flux contribution from the surrounding H\,\textsc{ii} region. We find that unlensed population III stars are beyond the reach of {\it JWST}, and that even lensed population III stars will be extremely difficult to detect. However, the main problem with the latter approach is not necessarily that the lensed stars are too faint, but that their surface number densities are too low. To detect even one 60 M$_{\odot}$ population III star when pointing {\it JWST} through the galaxy cluster MACS J0717.5+3745, the lensing cluster with the largest Einstein radius detected so far, the cosmic star formation rate of population III stars would need to be approximately an order of magnitude higher than predicted by the most optimistic current models.

\end{abstract}


\section{Introduction}
\label{introduction}

Both theoretical arguments and numerical simulations \citep[e.g.][]{2004ARA&A..42...79B}, strongly support the notion that population III stars were very massive, significantly more so than the population I and population II stars that formed later on. When stars form in metal-enriched gas, the Jeans mass is lower, which leads to fragmentation and thus lower masses. Lacking metals, a chemically unenriched cloud does not fragment in the same way and the characteristic stellar mass may therefore be higher. It has been argued that two different classes may have existed: population III.1 and population III.2. Population III.1 stars formed in dark matter minihalos of mass 10$^5$--10$^6$ M$_{\odot}$ at $z\approx$ 10--30. Early research on population III star formation \citep{2007ApJ...654...66O, yoshidaomukaihernquist2008, 2009Natur.459...49B}, indicated that only one star, with a very high average mass of $\sim$ 100 M$_{\odot}$ was produced in every minihalo. This was the result of UV radiation (in the Lyman-Werner band) produced by the first massive star destroying the molecular hydrogen in the parent cloud, preventing further cooling and star formation. Later research on the same topic \citep[e.g][]{2009Sci...323..754K, 2009Sci...325..601T, 2010MNRAS.403...45S, 2011ApJ...731L..38P, 2011ApJ...727..110C, 2011ApJ...737...75G, 2012MNRAS.422..290S} indicates a more complex scenario with more fragmentation, resulting in a binary system or small cluster of stars, with the most massive star reaching $\approx$~50~M$_{\odot}$. \citet{2008ASPC..393..275S} also suggest that cosmic rays from the supernovae resulting from the first massive population III stars could significantly influence subsequent star formation in minihalos, possibly yielding a characteristic stellar mass of $\sim$~10~M$_{\odot}$. It is also plausible that protostellar feedback will halt the mass accretion, thereby limiting the stellar mass to a few times 10~M$_{\odot}$ \citep{2011Sci...334.1250H}.

There are also models indicating late, $2 < z < 7$, population III star formation \citep{2007MNRAS.382..945T, 2010MNRAS.404.1425J}. These late population III stars form in regions that are pristine due to highly inhomohegeneous metal enrichment. These low-metallicity regions are usually located in the vicinity of more massive ($\sim 10^8-10^{11}$~M$_{\odot}$) dark matter halos containing metal enriched galaxies. In \citet{2011MNRAS.411.2336I}, the authors examine LAEs at $z=3.1$ with signatures they argue are hard to interpret as other than objects containing population III stars, (but see \citet{2012MNRAS.tmp..127F} for a different view).

Population III.2 stars are thought to form mostly due to cooling by adiabatic expansion or HD cooling, usually at a lower redshift than population III.1 stars, and are thought to be building blocks for early galaxies. Since we focus on pre-galactic population III star formation we will not investigate these any further.

\citet{2001ApJ...552..464B} have investigated the spectral energy distribution of a primordial star. \citet{2006SSRv..123..485G} and \citet{2009MNRAS.399..639G} used these results to show that isolated population III stars are too faint for detection with the {\it JWST}. However, the flux boost due to gravitational lensing has not previously been considered in detail, except for a preliminary version of the study we present here \citep{2010crf..work...26R}.

Apart from direct detection through gravitational lensing, there are other ways to probe the properties of population III stars. The cumulative indirect signatures in the global radio background produced by bremsstrahlung or the 21-cm hydrogen emission also represent interesting options. Bremsstrahlung is free-free emission originating in the H\,\textsc{ii} region in its active and relic states. When the population III star has died the almost fully ionized H\,\textsc{ii} region around it begins to recombine. This makes it a very bright source of hydrogen 21-cm radiation \citep{2009MNRAS.395..777T}. \citet{2009MNRAS.399..639G} find that bremsstrahlung is very unlikely to be detected in the near future but the cumulative 21-cm radiation might be strong enough to be within reach of the upcoming Square Kilometer Array. Direct observation of pair instability supernovae (PISN) could also be used to infer the existence of population III stars. PISNs are the predicted final fate of non-rotating stars in the mass range 140--260~M$_{\odot}$ \citep{1967PhRvL..18..379B, 1968Ap&SS...2...96F, 2012arXiv1211.4979W, 2012MNRAS.tmp..244D}, and are possible for lower mass stars if rotating \citep{2012ApJ...760..154C}. The initial mass function (IMF) describes the distribution of mass for a new star. Since PISN occur in a certain mass range observations of those could probe the IMF. Using a semi-analytic halo mass function approach, \citet{2012ApJ...755...72H} find that no more than $\sim$~0.2 population III PISN should be visible per {\it JWST} field of view at any one time, depending on the adopted model for stellar feedback. Rapidly rotating population III stars may also produce collapsar gamma-ray bursts, or GRBs \citep{2011A&A...533A..32D}. These are formed by metal poor, rotating stars more massive than $\sim$~40~M$_{\odot}$. This could be a highly energetic probe of population III stars, and of the high-mass tail of their IMF. Population III stars could also leave detectable imprints \citep{2006astro.ph.10943K} in the cosmic infrared background (CIB). The CIB is a repository of emissions from different sources at different redshifts, all redshifted into the infrared. The imprints from population III stars are already constrained by detection of emission at energies above 10~GeV \citep{2012MNRAS.420..800G}, as interaction between the radiation of population III stars and gamma rays create a gamma-ray optical depth.

As we will argue, a flux boost due to gravitational lensing by a foreground galaxy cluster will be necessary to directly detect population III stars. By adopting the properties of one of the more promising lensing clusters, we will calculate the SFR required to achieve $\sim$~1 detectable population III star per survey field.

\section{Modeling the spectra of Population III stars}
\label{sec:modellingspectrapopulationIIIstars}

We have used the properties of population III stars derived by \citet{2002A&A...382...28S} as the basis for our estimated broadband fluxes. Here, we adopt the main sequence properties of these stars and thereby neglect the fact that their ionizing fluxes decrease as they grow older. This results in a mild overestimate of the flux in the filters we are using. We have used realistic stellar atmospheres and have also taken into account the potential flux contribution from the surrounding H\,\textsc{ii} region. To model the stellar atmospheres, we have used the publicly available TLUSTY code, \citep{1995ApJ...439..875H}. This code computes 1D, non-LTE, plane-parallel stellar model atmospheres and spectra, given a certain input stellar composition, surface temperature and gravity. We set the chemical composition of our stars to the primordial mixture of H and He \citep{2009PhR...472....1I}.

For our models of population III stars, the luminosity and number of ionizing photons (important for the luminosity originating in the nebula around the stars) scale approximately linearly with stellar mass for $M$~$>$~25~M$_{\odot}$, but not below this limit. This means that the luminosity will be dominated by a few massive stars, one or possibly two of a few tens of solar masses \citep{2009Sci...323..754K, 2009Sci...325..601T, 2010MNRAS.403...45S, 2011ApJ...731L..38P, 2011ApJ...727..110C, 2011ApJ...737...75G, 2011Sci...334.1250H, 2012MNRAS.422..290S}. Their combined luminosity will correspond to a star with stellar mass slightly lower than the sum of their masses, due to the non-linearity of the luminosity as a function of stellar mass below 25~M$_{\odot}$. Smaller stars add very little compared to larger stars. For example, a 9~M$_{\odot}$ star has an energy output (bolometric and ionizing) that corresponds to less than 10\% of that of a 25~M$_{\odot}$ star. In this paper we will work with one population III star as a proxy for small clusters of this type. We select a 60~M$_{\odot}$ star as our default option because its output likely exceeds the output of a smaller cluster. As an example of a more extreme scenario, we will also present calculations for a stellar mass of 300~M$_{\odot}$; this is the lowest stellar mass that turns out to be readily detectable in our models. A 200~M$_{\odot}$ star, for example, is only detectable for a small subset of models, and only over a small redshift interval. Our proxy stars admittedly have lower lifetimes than the constituents of the cluster they are representing. This means an argument could be made for using longer lifetime in our calculation, thereby improving the prospects of detection. Because the lifetimes do not differ by much (a 40~M$_{\odot}$ and 60~M$_{\odot}$ star differ 10\% in lifetimes, as an example), and it would be difficult to make a more accurate estimate of the effect, we will adopt the lifetime of the proxy star as representative of the lifetime of a small cluster.

The observed flux from a population III star is probably strongly influenced by the H\,\textsc{ii} region surrounding it \citep{2004ApJ...613..631K}. To model this, we have used the publicly available photoionization code Cloudy \citep{1998PASP..110..761F}. Assuming spherically symmetric H\,\textsc{ii} regions, we use Cloudy to calculate the resulting spectrum. We predict both the nebular continuum and emission lines. This procedure results in an optimistically bright emission spectrum, as more realistic nebulae experience feedback effects that could give rise to holes in the H\,\textsc{ii} region. This exposes the star underneath and the spectrum emerging from the stellar atmosphere. The region could also break out of the gravitational well of the host halo if the feedback is strong enough \citep[e.g.][]{2004ApJ...613..631K, 2004ApJ...610...14W}. As described by \citet{2009MNRAS.399..639G}, this dilutes the nebular flux. This could expose more of the purely stellar continuum resulting in a different overall spectrum \citep{2011ApJ...740...13Z}.

For the low densities expected in the H\,\textsc{ii} regions around the stars we have considered, time dependent effects start to become important at late times when the density may become as low as 1~atom~cm$^{-3}$, or lower. As Cloudy is a steady-state code, we have tested the importance of time-dependence using the latest version of the time-dependent photoionization code described in \citet{1996ApJ...464..924L}, \citet{1999ApJ...511..389L,2007AIPC..937..102L} and \citet{2010ApJ...717.1140M}. For hydrogen and helium, a total of 56 levels for H\,\textsc{i--ii} and 72 levels for He\,\textsc{i--iii} are included, and all levels are treated time-dependently. We discuss the importance of time dependence for the emission from the H\,\textsc{ii} region in Section \ref{sec:discussion}.

At $z>6$ the Gunn-Peterson trough \citep{1965ApJ...142.1633G} affects each emitted spectrum, resulting in huge absorption of radiation with wavelength lower than the Lyman-$\alpha$ (Ly$\alpha$) line. To simulate this we set all flux at wavelengths shorter than Ly$\alpha$ to zero at $z>6$. For $z<6$ the situation is more complicated as the radiation passes through several clouds of neutral hydrogen. Each cloud absorb Ly$\alpha$ radiation producing an absorption line. The resulting absorption lines are called the Ly$\alpha$-forest. The exact nature of the Ly$\alpha$-forest is dependent on the line of sight with the average absorption increasing with redshift. We have used the \citet{1995ApJ...441...18M} model to modify the radiation below the Ly$\alpha$-line at $z<6$. Even though Ly$\alpha$ is the main absorbing mechanism the model also takes higher order Lyman-absorbers into account. We use the average absorption predicted by the model.

The very high effective temperature of a population III star ($\sim$~100,000~K) implies a huge hydrogen ionizing flux. This is converted into lower energy radiation but also into a substantial number of Ly$\alpha$ photons. This can potentially make the Ly$\alpha$ line the strongest rest frame emission line in the spectra of these population III stars. At the same time, this is a highly resonant line which means that it suffers from absorption in the neutral intergalactic medium (IGM). The Ly$\alpha$ photons must also escape the interstellar medium (ISM) where they scatter and might be destroyed by dust. Even so, if just a fraction of these photons manage to escape the ISM and IGM absorption they could significantly improve the prospects of detecting population III stars. The combined fraction of Ly$\alpha$ photons escaping both the ISM and the IGM will be denoted $f_\mathrm{Ly\alpha}$. Our default assumption is to set the Ly$\alpha$ line to zero ($f_\mathrm{Ly\alpha}=0$), but we will also consider two scenarios with strong escape fraction, setting $f_\mathrm{Ly\alpha}$ to 0.2 or 0.5.

\begin{figure}
    \begin{center}
        \includegraphics[width = 10 cm]{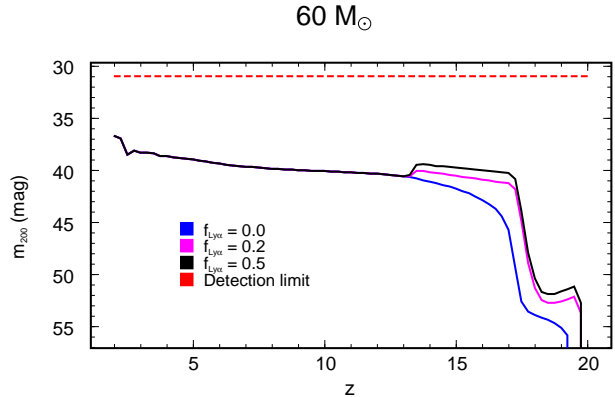}
        \caption{The {\it NIRCam/F200W} AB magnitude predicted for a 60~M$_{\odot}$ star as a function of redshift in an unlensed field. The blue line represents the magnitude with no Ly$\alpha$ radiation escaping, whereas the purple and black lines represent $f_\mathrm{Ly\alpha}=0.2$ and 0.5, respectively. The red dashed line indicates the detection limit (30.9~mag) considering $S/N=10$, using an exposure time of 100h. As seen, the unlensed flux of a 60~M$_{\odot}$ population III star lies $\gtrsim$~6 magnitudes below the {\it JWST} detection threshold.}
        \label{fig:magnitude}
    \end{center}
\end{figure}

\subsection{MACS J0717.5+3745}
\label{sec:macsj0717}

To assess the prospects for directly detecting a 60~M$_{\odot}$ population III star, we calculate the intrinsic {\it JWST} broadband flux of such a star as a function of redshift and compare it to the detection limit. In Figure~\ref{fig:magnitude}, we display the results for the {\it JWST/NIRCam F200W} filter. The sensitivity of {\it NIRCam} when using {\it F200W} is generally the most promising for detection of population III stars at the redshifts we are interested in. The brightest population III magnitude is 36.6 at a redshift of 2. For comparison, the $S/N=10$, 100h exposure time detection limit is 30.9, placing this population III star $\gtrsim$~6 magnitudes below the {\it JWST} detection limit.

As it appears impossible to detect a high redshift population III star in an unlensed field, we will in the following instead explore the prospects of detecting population III stars in gravitationally magnified regions of the sky. For this purpose we adopt MACS J0717.5+3745, a galaxy cluster at $z=0.546$, as our lensing cluster. This is the gravitational lens with largest angular Einstein radius known, and also one of the targets of the ongoing {\it HST} survey {\it CLASH}\footnote{{\it Cluster Lensing And Supernova survey} with {\it Hubble}} \citep{2012ApJS..199...25P}.

\citet{2009ApJ...707L.102Z} presented a detailed model of this particular gravitational lens. The unusual structure of the object, with an unrelaxed morphology and correspondingly shallow density profile, results in exceptionally large areas with high magnification. For instance, at $z>6$, the area over which background objects achieve a magnification higher than 10 is $\approx$~3.5~arcmin$^2$. The data we have for the magnification of MACS J0717.5+3745 cover $z>6$. For lower redshifts we have used the data for $z=6$, this should only have a small impact on our calculations as the magnified area changes slowly with redshift.

\section{The SFR required for JWST detection of population III stars in lensed fields}
\label{sec:JWSTbroadbandfluxes}

\begin{figure}
    \begin{center}
        \includegraphics[width = 10 cm]{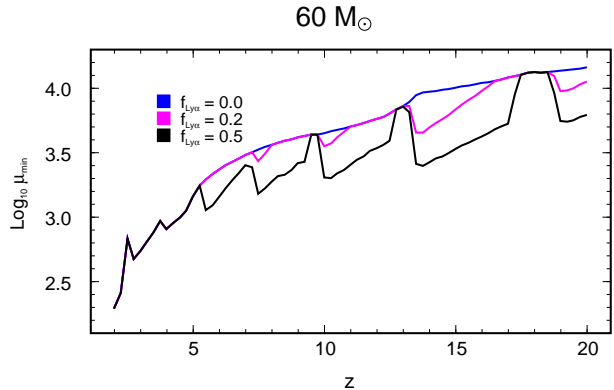}
        \caption{The minimum gravitational magnification $\mu_\mathrm{min}$ required to detect a 60~M$_{\odot}$ population III star as a function of redshift. The detection criterion is based on a $10 \sigma$ detection after an exposure time of 100h. For each redshift, the filter resulting in the lowest necessary magnification is selected in the calculation of $\mu_\mathrm{min}$. Clearly, direct detection without gravitational lensing is impossible and even when seen through a lensing cluster, a very high magnification ($\mu \gtrsim 200$, increasing steeply with redshift) is required. The blue line represents the lowest required $\mu$ with no Ly$\alpha$ radiation escaping, whereas the purple and black lines represent $f_\mathrm{Ly\alpha}=0.2$ and 0.5, respectively.}
    \label{fig:necessarymagnification}
    \end{center}
\end{figure}

Figure \ref{fig:necessarymagnification} shows the minimum magnification ($\mu_\mathrm{min}$) required for detection of one 60~M$_{\odot}$ star as a function of redshift in the J0717.5+3745 field. At each redshift the filter requiring the lowest magnification is used when calculating $\mu_\mathrm{min}$. We consider a generous exposure time of $t_\mathrm{exp}=100$~hours and use a detection limit of $S/N=10$ per pointing\footnote{Note that it is possible that it takes two pointings to cover the J0717.5+3745 caustic (high magnification area) with {\it JWST/NIRCam}.}. For a fixed detection limit, $S/N \propto \sqrt{t_\mathrm{exp}}$. This means, as an example, that all results hold for $S/N=2$ using $t_\mathrm{exp}=4$h and for $S/N=5$ using $t_\mathrm{exp}=25$h, as well as our default $S/N=10$ using the $t_\mathrm{exp}=100$h case. Included are also two different non-zero Ly$\alpha$ escape fractions: $f_\mathrm{Ly\alpha}$ = 0.2 and 0.5. A positive $f_\mathrm{Ly\alpha}$ facilitates detection at most redshifts. For redshifts where the Ly$\alpha$ line is located between filters the required magnification is of course unchanged in each filter. For a 60~M$_{\odot}$ star, magnifications $\gtrsim 200$ are required, as seen in Figure~\ref{fig:necessarymagnification}. Magnifications this large are probably not impossible, but will be attained only over very small areas in the lensing cluster \citep{2009ApJ...707L.102Z}.

Using the result from Figure~\ref{fig:necessarymagnification}, we calculate the total lensed area behind MACS J0717.5+3745 that displays the required magnification for detection in a unit redshift interval as a function of redshift. The corresponding co-moving cosmological volume is calculated using $H_0$=72~km~s$^{-1}$~Mpc$^{-1}$, $\Omega_\mathrm{m} = 0.27$ and $\Omega_{\Lambda}=0.73$. We have adopted the lifetimes of population III stars given in \citet{2002A&A...382...28S} for this calculation, and considered the star formation rate (SFR) constant throughout this time interval at each redshift. By combining the adopted typical mass of population III stars with their predicted lifetimes, we arrive at a lower limit (SFR$_{\mathrm{min}}$) on the SFR required to form at least one population III star within this volume.

We have explored three different star formation models from \citet{2009ApJ...694..879T}, the simulations from \citet{2012ApJ...745...50W} and \citet{2007MNRAS.382..945T}, to compare with our results for SFR$_{\mathrm{min}}$. The three \citet{2009ApJ...694..879T} models use H$_2$ cooling in minihalos to form the stars, the difference between them being how they handle the expected Lyman-Werner (L-W) radiation feedback. L-W radiation is emitted by hot stars and destroys H$_2$ through photodissociation, thereby prohibiting further population III star formation. The first, standard, model assumes full feedback from L-W which effectively prevents population III star formation at $z<13$. The other two models assume reduced and no L-W feedback. These scenarios correspond to situations where the L-W radiation is lowered for some reason. This could happen in the case where the ``isolated"  population III stars we consider in fact belong to small clusters  with many low-mass stars with low L-W fluxes, thereby decreasing the effective L-W flux per unit stellar mass. Instabilities in the ionization fronts resulting in self shielding could also hamper L-W feedback \citep{2008ApJ...673..664W}.

\citet{2012ApJ...745...50W} use an adaptive mesh radiation hydrodynamics simulation to model star formation. In their models, the ionizing and dissociating radiation from nearby population III stars and galaxies proves to be unable to halt population III star formation. Instead, it only suppresses it, resulting in a nearly constant SFR around $5 \times 10^{-5}$~M$_{\odot}$~yr$^{-1}$~Mpc$^{-3}$ throughout the redshift interval $z=7$--$18$.

To compare our results to lower redshift population III star formation, possible through inhomogeneous metal enrichment, we use the standard model in \citet{2007MNRAS.382..945T}. Their code uses smoothed  particle hydrodynamics to simulate the development of star formation. The suppression of star formation at $z<6$ is caused by the IGM photoheating due to reionization. The collapse of low-mass structures is inhibited as the photoheating increases the Jeans scale.

\begin{figure}
    \begin{center}
        \includegraphics[width = 10 cm]{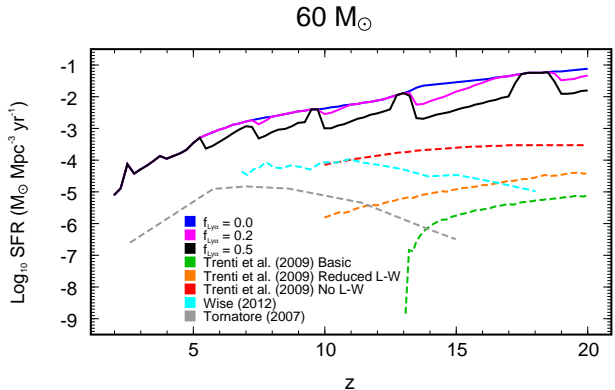}
        \caption{The minimum star formation rate SFR$_{\mathrm{min}}$ required for approximately one 60~M$_{\odot}$ star to be detected when {\it JWST} is pointed through the galaxy cluster MACS J0717.5+3745. The detection criterion is based on a $10 \sigma$ detection after an exposure time of 100h. The blue solid line represents the SFR$_{\mathrm{min}}$ with no Ly$\alpha$ radiation escaping, whereas the purple and black solid lines represent $f_\mathrm{Ly\alpha}=0.2$ and 0.5, respectively. Different SFR models are also included for comparison, using dashed lines. The SFR models by \citet{2009ApJ...694..879T} are represented by the green, orange, and red lines. They represent different levels of L-W radiation feedback, a feedback prohibiting further star formation. The three levels are basic (green), meaning full L-W feedback, reduced (orange), and no L-W (red). The models by \citet{2012ApJ...745...50W} and \citet{2007MNRAS.382..945T} are included in cyan and grey respectively. The figure shows that there is a gap of an order of magnitude between the most optimistic simulation and the required SFR for a detection, even when a very high Ly$\alpha$ escape fraction is assumed.}
    \label{fig:requiredSFR60}
    \end{center}
\end{figure}

\begin{figure}
    \begin{center}
        \includegraphics[width = 10 cm]{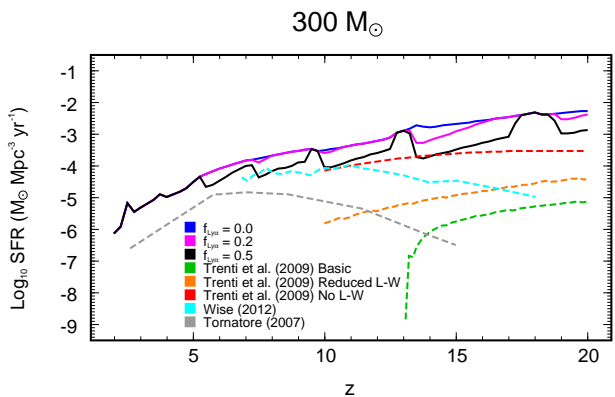}
        \caption{Same as Figure~\ref{fig:requiredSFR60}, but for a 300~M$_{\odot}$ star. With this fairly extreme stellar mass, a detection would be possible with the most optimistic \citet{2009ApJ...694..879T} simulation and the \citet{2012ApJ...745...50W} model, but only when a very high Ly$\alpha$ escape fraction is assumed. The mass 300~M$_{\odot}$ is selected for the star to be readily detectable with the star formation models we compare to. A 200~M$_{\odot}$ star, for example, is not detectable with the \citet{2009ApJ...694..879T} models even though it still is detectable at some redshifts when employing the \citet{2012ApJ...745...50W} model.}
    \label{fig:requiredSFR300}
    \end{center}
\end{figure}

In Figure~\ref{fig:requiredSFR60}, we plot SFR$_{\mathrm{min}}$ against redshift for the case of a 60~M$_{\odot}$ star. This graph includes the two cases with non-zero $f_\mathrm{Ly\alpha}$ and our comparison SFR models. There is a gap of an order of a magnitude between the most optimistic simulations and SFR$_{\mathrm{min}}$, even when considering a very high Ly$\alpha$ escape fraction. Thus, our default scenario is undetectable. The corresponding requirements for a 300~M$_{\odot}$ star are shown in Figure~\ref{fig:requiredSFR300}. In this case, a detection is possible if considering one of the most optimistic SFR predictions and a very high Ly$\alpha$ escape fraction.

\begin{figure}
    \begin{center}
        \includegraphics[width = 10 cm]{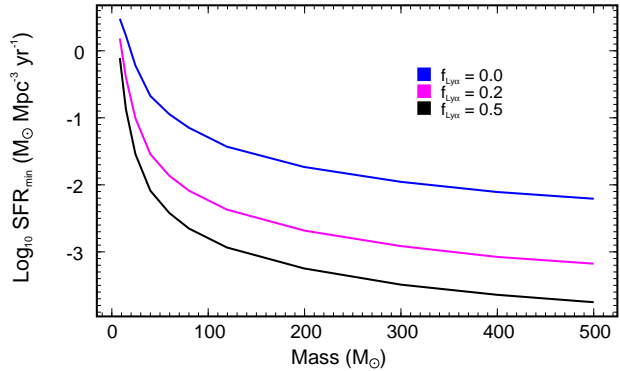}
        \caption{The SFR$_{\mathrm{min}}$ in the MACS J0717.5+3745 field, as a function of the population III mass (assuming all such stars to have the same mass) at $z=15$. Imaging in the {\it NIRCam/F200W} filter is assumed. Since this graph uses a single filter, it is not necessarily the filter resulting in the lowest SFR$_{\mathrm{min}}$. This means that the values in this graph differ in some instances from the values in Figure~\ref{fig:requiredSFR60} and \ref{fig:requiredSFR300}, where the filter resulting in the lowest SFR$_{\mathrm{min}}$ is always selected. The detection criterion is based on a $10 \sigma$ detection after an exposure time of 100h. The blue line represents the SFR$_{\mathrm{min}}$ with no Ly$\alpha$ radiation escaping, whereas the purple and black lines represent $f_\mathrm{Ly\alpha}=0.2$ and 0.5, respectively. A higher population III stellar mass results in a lower required SFR. This indicates that a top heavy IMF improves the prospects of detecting high-redshift population III stars.}
    \label{fig:sfrrequiredfrommass}
    \end{center}
\end{figure}

The IMF of population III stars is essentially unknown. Simulations indicate a high typical mass but the exact distribution is important for the prospects of detection. In Figure~\ref{fig:sfrrequiredfrommass} SFR$_{\mathrm{min}}$ is plotted as a function of the typical mass of population III stars at $z=15$ in the filter {\it F200W}. We selected this redshift and filter passband in order to minimize SFR$_{\mathrm{min}}$, while simultaneously maximizing the potential impact of the Ly$\alpha$ emission line. To produce a given number of stars, higher stellar masses imply a higher SFR, in order to supply the additional mass. More massive stars also live shorter lives, further increasing SFR$_{\mathrm{min}}$. On the other hand, a higher mass implies a higher luminosity. This means that a larger cosmological volume in a unit redshift has sufficient magnification for observation, implying a lower SFR$_{\mathrm{min}}$. It turns out that the effect of the luminosity trumps the mass and lifetime effects. Hence, a larger mass implies a lower required SFR, as can be seen in Figure \ref{fig:sfrrequiredfrommass}. Thus a top-heavy IMF improves the prospects of detecting population III stars.

So far, we have concentrated on the prospect of detecting population III stars in a specific redshift bin by selecting the filter resulting in the lowest SFR$_{\mathrm{min}}$. Broadband imaging surveys will in reality probe a wide range of redshifts simultaneously. For the purposes of comparison, here we consider the redshift interval 7--15, which roughly corresponds to the \citet{2012ApJ...745...50W} model. For computational purposes, we kept the SFR constant in the redshift interval. We arrive at a slightly lower cosmic SFR$_{\mathrm{min}}$ compared to the redshift bin approach. This is more or less consistent with the \citet{2012ApJ...745...50W} model, but in reality could of course result in a more complicated SFR. Table~\ref{tab:cosmicSFR} displays the result for the different {\it JWST} filters in the {\it NIRCam} and {\it MIRI} instruments. Comparing to the lowest values in Figure~\ref{fig:requiredSFR60}, we see that larger redshift windows do not significantly alter our conclusions. This conclusion follows mostly from the IGM absorption. In the filters that return the lowest SFR$_{\mathrm{min}}$ (which occurs at low redshifts), the radiation coming from higher redshifts is mostly absorbed by the IGM. Hence, the final SFR$_{\mathrm{min}}$ in those filters is dominated by a small interval at low redshift.

\begin{table*}
 \centering
 \begin{minipage}{140mm}
  \caption{The minimum log$_{10}$~SFR$_{\mathrm{min}}$~(M$_{\odot}$~Mpc$^{-3}$~yr$^{-1}$) required for detection of one 60~M$_{\odot}$ population III star {\it anywhere} within the redshift interval $7<z<15$, as a function of the filter used. A constant SFR over the relevant redshift interval is assumed. The detection criterion is based on a $10 \sigma$ detection after an exposure time of 100~h. Three cases are considered: no escape through ISM and IGM of Ly$\alpha$ photons and the cases $f_\mathrm{Ly\alpha}=0.2$ and 0.5. The best filter choices for each $f_\mathrm{Ly\alpha}$ have been marked in boldface. The radiation through the lower wavelength filters {\it F070W} and {\it F090W} are to the most part absorbed by the Gunn-Peterson trough and so are excluded. We do not include {\it MIRI} for the longer wavelength filters, as its sensitivity is too low.}
  \begin{tabular}{@{}llrrrr@{}}
            Filter & Instrument & $\lambda$ ($\mu$m)  & f$_{Ly\alpha}$ = 0.0  & f$_{Ly\alpha}$ = 0.2 & f$_{Ly\alpha}$ = 0.5\\
             &  &  &  &  & \\
            {\it F115W} & {\it NIRCam} & 1.15 & $-$2.38 & $-$2.95 & \textbf{$-$3.44}\\
            {\it F150W} & {\it NIRCam} & 1.50 & $-$2.99 & $-$3.14 & $-$3.41\\
            {\it F200W} & {\it NIRCam} & 1.99 & \textbf{$-$3.23} & \textbf{$-$3.27} & $-$3.36\\
            {\it F277W} & {\it NIRCam} & 2.79 & $-$3.04 & $-$3.04 & $-$3.04\\
            {\it F356W} & {\it NIRCam} & 3.56 & $-$3.03 & $-$3.03 & $-$3.03\\
            {\it F444W} & {\it NIRCam} & 4.44 & $-$2.64 & $-$2.64 & $-$2.64\\
            {\it F560W} & {\it MIRI} & 5.63 & $-$1.19 & $-$1.19 & $-$1.19\\
            {\it F770W} & {\it MIRI} & 7.65 & $-$0.72 & $-$0.72 & $-$0.72\\
            \label{tab:cosmicSFR}
\end{tabular}
\end{minipage}
\end{table*}

\section{Discussion}
\label{sec:discussion}

Of the many scenarios explored in Section~\ref{sec:JWSTbroadbandfluxes}, the only one that would make isolated population III stars potentially detectable with {\it JWST} is:

\begin{itemize}

\item Very massive ($\gtrsim 300$~M$_{\odot}$) population III stars.
\item The stars are formed in the limiting case of either no L-W feedback \citep{2009ApJ...694..879T} or in the prolonged population III star formation case predicted by \citet{2012ApJ...745...50W}. 
\item The Lyman-alpha escape fraction would need to be very high ($f_\mathrm{Ly\alpha} \approx 0.5$).
\item The gravitational lens contains sufficiently large regions with very high magnifications ($\gtrsim$~1000).
\item The escape of Lyman continuum (i.e. hydrogen-ionizing) photons from the nebula surrounding the population III stars is very small.

\end{itemize}

Hence, the prospects for detecting isolated population III stars in the foreseeable future appear bleak. However, contrary to previous claims \citep{2006SSRv..123..485G, 2009MNRAS.399..639G}, the problem is not necessarily that population III stars are too faint for detection, as the magnification of a suitably chosen lensing cluster may lift them above the {\it JWST} detection threshold. The main obstacle is instead that the surface number densities for sufficiently massive population III stars to be found in a sufficiently lensed field is likely to be too low.

Even so, given the scenario that we have a detection of a population III star we still have to identify it as such. One plausible mean of doing this suggested in the literature \citep{2001ApJ...553...73O, 2001ApJ...550L...1T} is to identify the He\,\textsc{ii}~1640~\AA{} line. Population III stars are predicted to have a significantly harder spectrum than even low metallicity stars. This means significantly more He are ionized in relation to H ionized. The result is that the ratio of the He\,\textsc{ii}~1640~\AA{} line to a non-resonant hydrogen line such as H$\alpha$ should be significantly larger for population III stars compared to that from even low metallicity stars.

It has also been argued that supersonic streaming velocities between dark matter and gas could delay population III star formation, \citep{2011ApJ...736..147G, 2011MNRAS.412L..40M}. The delay is substantial, $\Delta z \sim 4$, but the large shift in mass also means the number density of population III star-forming minihalos is reduced by up to an order of magnitude. There is criticism of the importance of the effect \citep{2011ApJ...730L...1S}, but if significant, it would of course impact the results presented here. The combined effect of fewer stars at lower redshift needs careful examination in future work.

\begin{figure}
    \begin{center}
        \includegraphics[width = 10 cm]{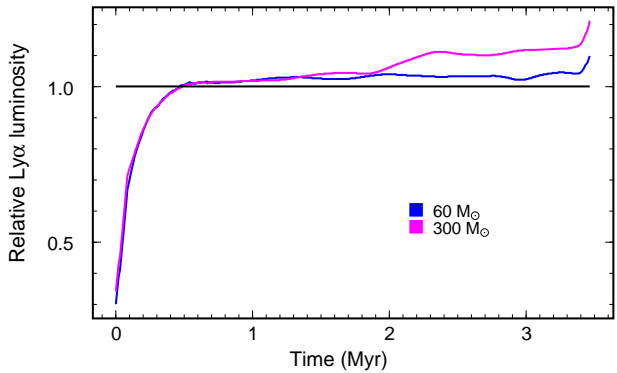}
        \caption{The flux in the Lyman alpha line from a $n_\mathrm{H} = 1$~cm$^{-3}$ nebula divided by the flux in the Lyman alpha line from a $n_\mathrm{H} = 100$~cm$^{-3}$ nebula, as a function of time. We have assumed that the stars suddenly switch on and shine for about 3.4~Myrs with a constant luminosity and spectrum. For the reference nebula with $n_\mathrm{H} = 100$~cm$^{-3}$ the ionization quickly goes toward the steady-state solution. The blue line represents a 60~M$_{\odot}$ star, whereas the purple line represents a 300~M$_{\odot}$ star. For $n_\mathrm{H} = 1$~cm$^{-3}$ it takes $\sim 0.4$~Myrs to reach the steady-state level. A 60~M$_{\odot}$ star's nebula subsequently emits $\sim$~5~\% more flux than in steady state and a 300~M$_{\odot}$ star's nebula up to $\sim$~15~\% more flux than in steady state. The upturn seen for both stellar models after 3.4~Myrs is an effect of time dependence in the low density (1~cm$^{-3}$) case. For both densities, the Ly-alpha emission starts to decay when the star is switched off, but the decay is about two orders of magnitude faster for 100~cm$^{-3}$ than for 1~cm$^{-3}$.}
    \label{fig:lyatimedependence}
    \end{center}
\end{figure}

As mentioned in Section~\ref{sec:modellingspectrapopulationIIIstars}, the low densities expected in the H\,\textsc{ii} regions could make time dependence important, in particular when the time scales of ionization/recombination and cooling/heating become comparable to the lifetimes of the stars. We have tested this using two scenarios for the surrounding H\,\textsc{ii} regions, one with 1 and another with 100 hydrogen atoms~cm$^{-3}$. We set the He/H number ratio to 0.09, and the abundances of other elements to zero. We show the relative luminosity of Ly$\alpha$ for the two densities and for both the 60~M$_{\odot}$ and 300~M$_{\odot}$ stars in Figure~\ref{fig:lyatimedependence}. We have assumed that the stars suddenly switch on and shine for about 3.4~Myrs with a constant luminosity and spectrum. For $n_\mathrm{H} = 100$~cm$^{-3}$, the ionization quickly goes toward the steady-state solution, whereas for $n_\mathrm{H} = 1$~cm$^{-3}$ it takes $\sim 0.4$~Myrs to reach the steady-state level. However, even after this level has been reached, there is still some slow photoionization going on in the outer parts of the H\,\textsc{ii} region. The temperature in this region also overshoots somewhat compared to the steady-state solution before steady state has been attained, which produces slightly enhanced collisional excitation of Ly$\alpha$. This effect is larger for the 300~M$_{\odot}$ star with its more extended region of partially ionized hydrogen outside the Str\"omgren radius. The net effect of time dependent versus steady-state models is, as can be seen from Figure~\ref{fig:lyatimedependence}, not more than $\sim 10-15$~\%. For densities below $n_\mathrm{H} = 1$~cm$^{-3}$, the deviations from steady-state will obviously become larger. However, the H\,\textsc{ii} region is expected to be evolving from high to low density with time \citep[e.g.][]{2004ApJ...610...14W}. This probably makes the early delay of Ly$\alpha$ in Figure~\ref{fig:lyatimedependence} exaggerated as, at this time, the H\,\textsc{ii} region is dense. At low densities the emission from the H\,\textsc{ii} region will fall off on a time scale comparable to the lifetime of the star. For such low densities, the H\,\textsc{ii} region may however no longer even be ionization bounded. The uncertainty in our models due to time dependence is in any case much smaller than due to other assumptions.

We have calculated the required magnification for detection of population III stars using realistic models for the atmosphere and the H\,\textsc{ii} region. However, our results are admittedly based on a number of simplifications, most of them made in the direction that would improve the prospects of detection. The spherically symmetric H\,\textsc{ii} region that we have assumed likely overestimates the flux, as there probably will be low-density channels in the nebula through which ionizing radiation would be able to escape into the intergalactic medium. We have also assumed a spatially homogeneous distribution of minihalos. In reality clustering could impact the detection probability in a negative way. All in all, even though not impossible, the prospect of detecting population III stars (isolated, or in small clusters) with {\it JWST} appears bleak at best.

Larger numbers of population III stars could potentially form in chemically pristine (10$^7$--10$^8$ M$_{\odot}$) atomic-cooling halos at $z<15$ \citep[e.g.][]{2012MNRAS.426.1159S}. These ``population III galaxies" would by comparison be much easier to detect and also identify \citep{2009MNRAS.399...37J, 2011ApJ...731...54P, 2011ApJ...740...13Z}. The detectability of such objects hinge more on the total star forming mass than on stellar masses and nebular properties, resulting in less extreme lensing magnification requirements \citep{2012arXiv1204.0517Z}. The merits of investigating the prospects for detecting population III stars in minihalos is that they possibly fill an important role in the development of the universe, providing the first metal-enrichment prior to the first galaxy formations \citep{2010ApJ...716..510G}.

More speculative and exotic theories involving population III stars or derivatives of them also exist, potentially improving detection prospects. Fast accretion combined with rapid rotation could potentially form super massive population III stars \citep{2012ApJ...756...93H}. These would be extremely massive, up to 10$^5$--10$^6$~M$_{\odot}$, and have lower temperature, $\sim$~5000~K. A black hole could also form by collapse in the center of the star powering it through emission by accretion, a so-called quasi-star \citep{2010MNRAS.402..673B}. Some significant fraction of population III stars could also be ``dark stars'', which would make the detection prospects slightly better (Zackrisson et al. 2010a)\nocite{2010ApJ...717..257Z}. There are also theories where dark stars accrete matter reaching masses of up to 10$^7$~M$_{\odot}$, which would improve prospects for observation significantly (\citealt{2010ApJ...716.1397F}; Zackrisson et al. 2010b\nocite{2010MNRAS.407L..74Z}; \citealt{2012MNRAS.422.2164I}).

\section{Summary}

In this paper, we have investigated the prospects of detecting high-redshift population III stars using the {\it JWST}. While isolated population III stars or small clusters of population III stars at $z > 2$ will be intrinsically too faint for detection, a foreground galaxy cluster acting as a gravitational lens could lift them above the detection threshold. We found that the detectability of single stars depends strongly on stellar mass, with more massive stars more easily detectable. With gravitational lensing we find for a 60~M$_{\odot}$ star at $z \sim 7.5$ a minimum Log$_{10}$~SFR of -3.4, considering a very high Ly$\alpha$ escape fraction of 0.5. For a 300~M$_{\odot}$ star we find a corresponding minimum Log$_{10}$~SFR of -4.4. Since our comparison simulation model at that redshift indicates a  Log$_{10}$~SFR of -4.3 this would actually imply a detection of $\sim 1$ population III star. So it turns out that unless the typical mass of population III stars approaches 300~M$_{\odot}$, their surface number density will be too low for detection.

\section*{Acknowledgments}
\label{sec:acknowledgement}

E.Z. and C-E.R. acknowledge funding from the Swedish National Space Board. C-E.R. also acknowledge funding from the Royal Swedish Academy of Sciences. E.Z. also acknowledge funding from the Swedish Research Council. P.L. is grateful for financial support from the Swedish Research Council. P.S. is supported by the Lorne Trottier Chair in Astrophysics and an Institute for Particle Physics Theory Fellowship.

\bibliographystyle{References}
\bibliography{References}

\end{document}